\documentclass[sigconf, nonacm]{acmart}
\newcommand\vldbavailabilityurl{https://doc-master.github.io/}
\newcommand\vldbpagestyle{plain}

\usepackage{booktabs}      % professional tables
\usepackage{multirow}      % multi-row table cells
\usepackage{balance}       % balance columns on last page
\usepackage{listings}      % code listings
\usepackage{xcolor}        % colors
\usepackage{xspace}        % smart spacing after macros
\usepackage{microtype}     % better typography
\usepackage{graphicx}      % figures
\usepackage{subcaption}    % subfigures
\usepackage{amsmath}       % math
\usepackage{enumitem}      % compact lists
\usepackage{pifont}        % \ding{51} checkmark, \ding{55} cross
\usepackage{etoolbox}      % \ifdefempty

\newcommand{\sys}{\textsc{DocMaster}\xspace}  % system name
\newcommand{\eg}{\textit{e.g.,}\xspace}

\newcommand{\etal}{\textit{et al.}\xspace}

\title{\sys: A Hierarchical Structure-Aware System for Document Analysis}
\subtitle{}

\author{Ziqi Chen}
\affiliation{%
  \institution{The Chinese University of Hong Kong, Shenzhen}
  \city{Shenzhen}
  \country{China}
}
\email{ziqichen1@link.cuhk.edu.cn}

\author{Yingli Zhou}
\affiliation{%
  \institution{The Chinese University of Hong Kong, Shenzhen}
  \city{Shenzhen}
  \country{China}
}
\email{yinglizhou@link.cuhk.edu.cn}

\author{Fangyuan Zhang}
\affiliation{%
  \institution{The Chinese University of Hong Kong}
  \city{Hong Kong}
  \country{China}
}
\email{zzzzzfy@link.cuhk.edu.hk}

\author{Quanqing Xu}
\affiliation{%
  \institution{OceanBase, AntGroup}
  \city{Hangzhou}
  \country{China}
}
\email{xuquanqing.xqq@oceanbase.com}

\author{Chuanhui Yang}
\affiliation{%
  \institution{OceanBase, AntGroup}
  \city{Hangzhou}
  \country{China}
}
\email{rizhao.ych@oceanbase.com}

\author{Yixiang Fang}
\affiliation{%
  \institution{The Chinese University of Hong Kong, Shenzhen}
  \city{Shenzhen}
  \country{China}
}
\email{fangyixiang@cuhk.edu.cn}

\begin{abstract}
Leveraging large language models (LLMs) to analyze complex documents --- such as academic papers, technical manuals, and financial reports --- has emerged as a mainstream and critical task in both research and industry.
In practice, users must first filter relevant documents from large collections and then conduct in-depth analysis (\eg question answering) over the selected subset, yet existing systems flatten documents into plain-text chunks, discarding the rich hierarchical structures (sections, tables, figures, equations) and degrading downstream performance.
We present \sys, a hierarchical structure-aware document analysis system.
\sys parses documents into hierarchical document trees preserving original layouts and constructs a structure-aware semantic index that enables accurate document filtering and in-depth analysis.
We demonstrate \sys through an interactive web interface that enables users to upload document collections, construct tree-based and multi-view semantic indices, filter relevant documents via natural-language conditions, and perform follow-up question answering over the filtered results.
The source code, data, and demo are available at \url{\vldbavailabilityurl}.
\end{abstract}

\begin{document}

\pagestyle{\vldbpagestyle}

\maketitle
% ============================================================
\section{Introduction}
\label{sec:intro}

\begin{figure}[t]
  \centering
    \includegraphics[page=4,
                    trim=3cm 2cm 3cm 2cm,
                    clip,
                    width=1.0\linewidth]{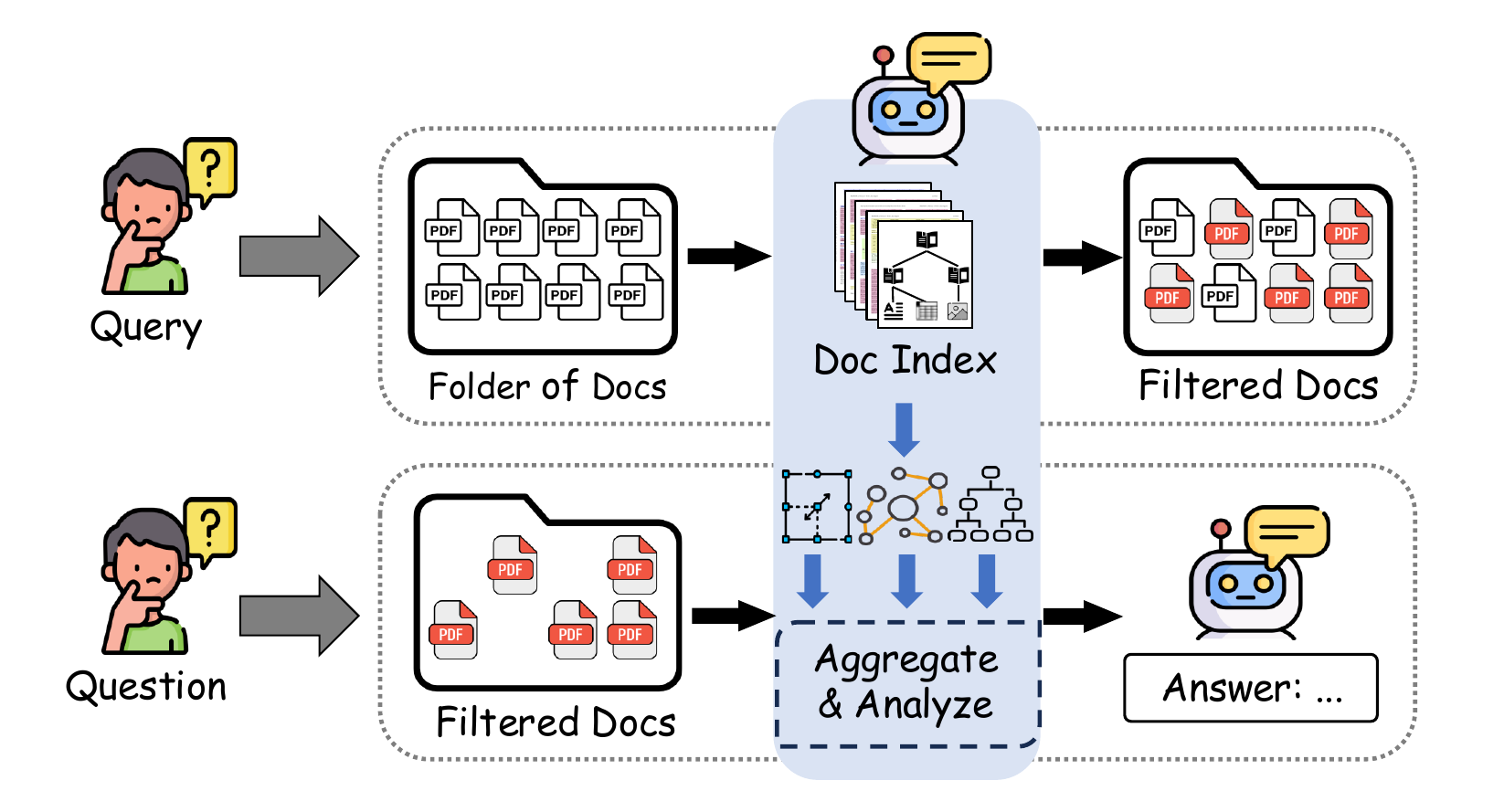}
  \caption{The overall workflow of document analysis.}
  \Description{Workflow showing document filtering followed by in-depth analysis over selected documents.}
  \label{fig:workflow}
\end{figure}

Leveraging large language models (LLMs) to analyze complex documents --- such as academic papers, technical manuals, and financial reports --- has emerged as a mainstream and critical task in both research and industry.
These documents encode information within rich hierarchical structures ---sections, subsections, tables, figures, and equations --- that must be understood for effective analysis.
In practice, users must first filter relevant documents from large collections and then conduct in-depth analysis (\eg question answering) over the selected subset, as shown in Figure \ref{fig:workflow}.
For example, a researcher surveying AI papers may need to find those that ``propose a retrieval-augmented method and evaluate on open-domain QA benchmarks'' before extracting detailed insights.

% --- Figure 1: Overall Pipeline ---

However, existing document analysis systems~\cite{lewis2020retrieval,Liu_LlamaIndex_2022,chase2022langchain} flatten documents into plain-text chunks, discarding the rich hierarchical structures and degrading downstream performance.
This structure-agnostic treatment introduces three key challenges.
\textbf{(C1) Hierarchy preservation.}
Flattening sections, subsections, and heterogeneous elements (text, tables, figures, equations) into flat chunks destroys structural relationships, causing systems to return context-poor fragments that hinder accurate filtering and answering.
\textbf{(C2) Cross-section semantic indexing.}
Implicit semantic relationships span across sections---\eg a method in Section~3 may be closely tied to an evaluation metric in Section~5.
Capturing such cross-section links requires indices that go beyond simple vector similarity.
\textbf{(C3) Cross-section evidence aggregation.}
A single filter condition may require evidence scattered across distant sections, yet existing systems retrieve chunks independently without aggregating cross-section evidence.

% --- Figure 2: System Layout Screenshot with Comments ---
\begin{figure*}[t]
  \centering
  \includegraphics[page=7,
                    width=0.9\linewidth]
                    {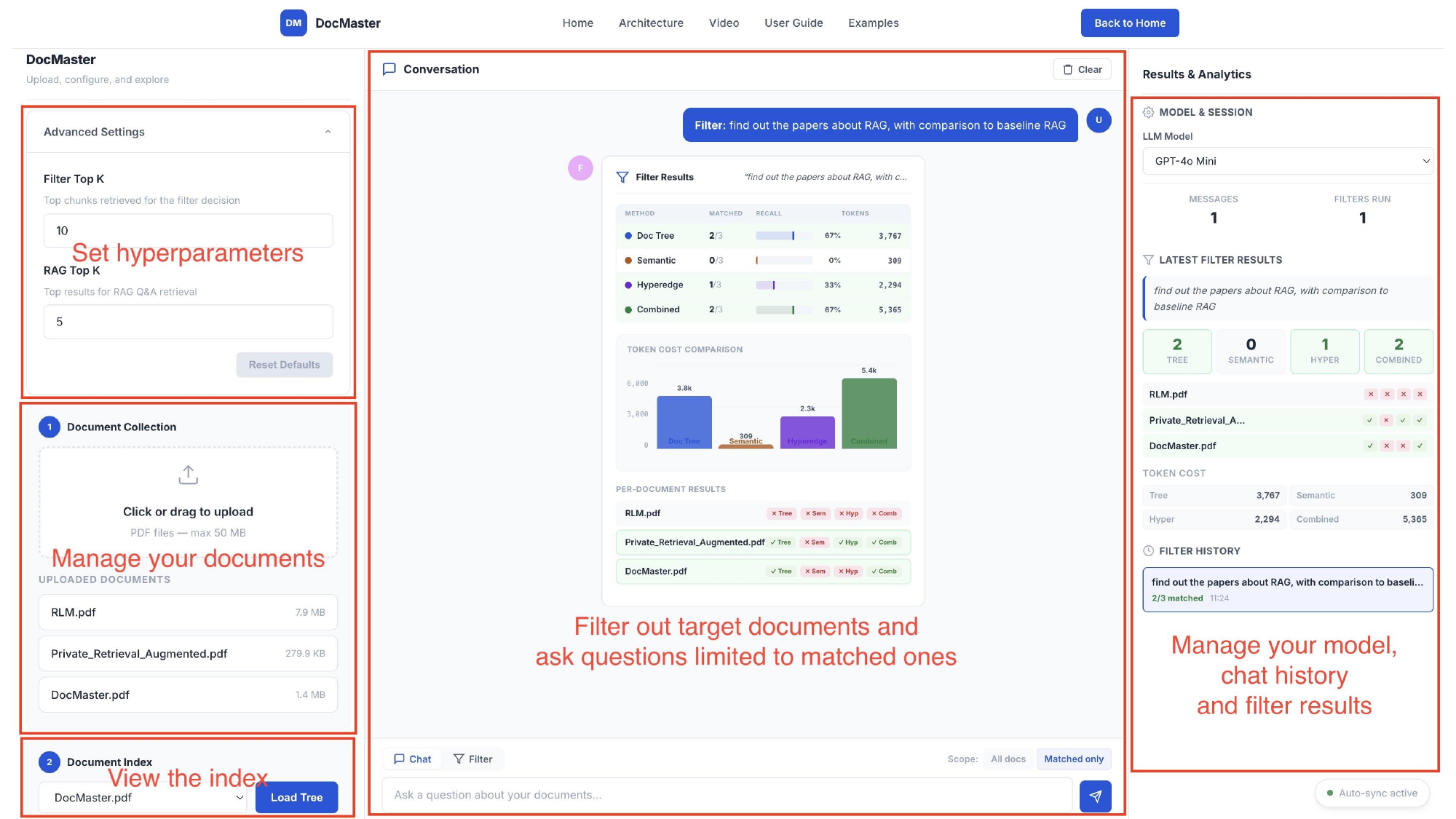}   
  \caption{The \sys web interface: users can issue natural-language filter queries, explore the document-tree index, tune hyperparameters, and compare filtering results side by side.}
  \Description{Screenshot of the DocMaster web interface with query, index exploration, parameter controls, and comparison panels.}
  \label{fig:system}
\end{figure*}

To address these challenges, we present \sys, a hierarchical structure-aware document analysis system.
\sys first parses documents into hierarchical document trees, preserving original layouts and constructs structure-aware semantic indices via LLM-guided constrained clustering (PC-KMeans) and cross-section hyper-edges.
It then performs tri-modal retrieval to accurately filter target documents from large collections, and finally leverages the filtered documents as grounded context for downstream question answering via retrieval-augmented generation (RAG)~\cite{gao2023retrieval}.

Our contributions are summarized as follows:
\begin{itemize}[leftmargin=*, nosep]
  \item We propose a hierarchical document representation that combines a structural document tree with a structure-aware semantic index, capturing both explicit hierarchy and implicit cross-section relationships (\textbf{C1, C2}).
  \item We design a tri-modal retrieval strategy integrating document-tree traversal, embedding-based semantic search, and hyper-edge matching for accurate document filtering (\textbf{C3}).
  \item We build an interactive web interface that enables users to upload document collections, construct tree-based and multi-view semantic indices, filter relevant documents via natural-language conditions, and perform follow-up question answering over the filtered results.
\end{itemize}

% ============================================================
\section{System Overview}
\label{sec:system}

% --- Figure 3: Demonstration Pipeline about RAG Paper ---
\begin{figure*}[t]
  \centering
  \includegraphics[page=5, 
                    trim=2cm 6cm 2cm 4cm,
                    clip,
                    width=0.9\linewidth]
                    {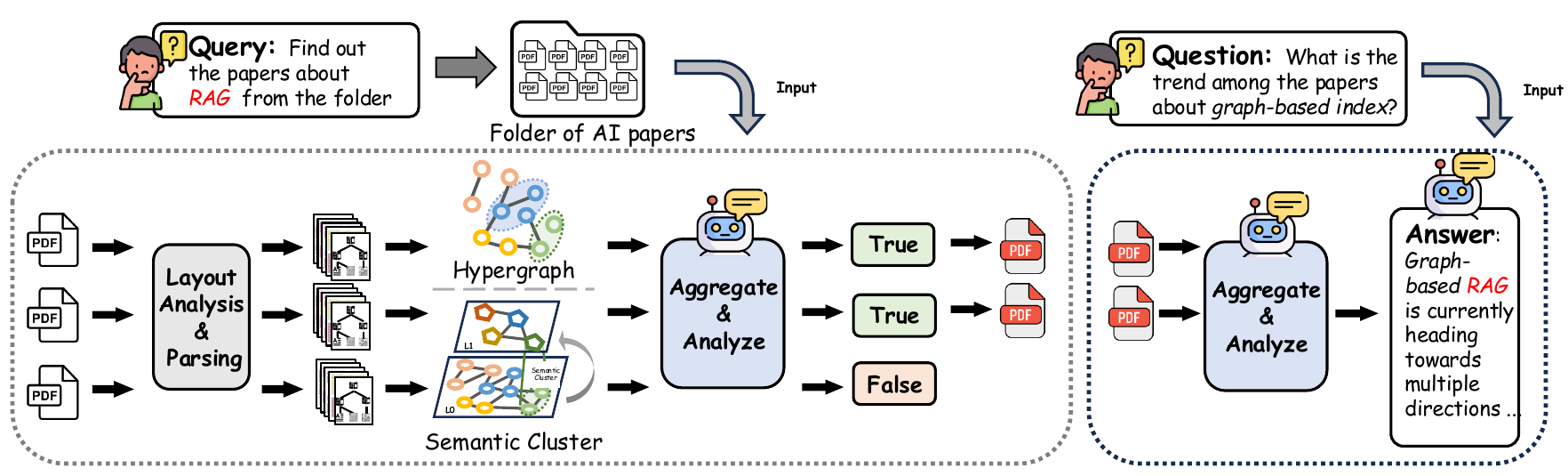}
  \caption{End-to-end example of \sys. The user uploads AI papers; each PDF is parsed into a hierarchical document tree, enriched with semantic indices (PC-KMeans clusters and hyper-edges), and queried through tri-modal retrieval for filtering. The user then asks follow-up questions about the filtered papers.}
  \Description{End-to-end demonstration pipeline from document upload to parsing, semantic indexing, filtering, and follow-up question answering.}
  \label{fig:usecase}
\end{figure*}

Figure~\ref{fig:workflow} illustrates the overall pipeline of \sys, which proceeds through four stages: (1)~document parsing, (2)~document tree construction, (3)~semantic index construction, and (4)~filtering and retrieval.

\subsection{Document Parsing}
\label{sec:parsing}

\sys employs MinerU~\cite{wang2024mineru} as its PDF extraction backend, converting documents into structured content while preserving layout information including text blocks, tables, figures, and equations.

\subsection{Document Tree Construction}
\label{sec:tree}

The parsed content is organized into a hierarchical document tree.
Each node is typed as one of \texttt{ROOT}, \texttt{TITLE}, \texttt{TEXT}, \texttt{IMAGE}, \texttt{TABLE}, or \texttt{EQUATION}.
Title nodes correspond to section and subsection headers, forming the internal structure of the tree, while content nodes (text, images, tables, equations) are attached as children under their enclosing titles.
Each node stores its raw content, a summary generated by GPT-4o mini, and a dense embedding computed by a sentence transformer (all-MiniLM-L6-v2~\cite{wang2020minilm}).
The tree is constructed bottom-up: leaf node summaries are generated first, then section summaries are recursively composed from their children.

\subsection{Semantic Index Construction}
\label{sec:semantic_index}

Beyond the structural document tree, \sys constructs a semantic index capturing implicit relationships across sections.
As illustrated in Figure~\ref{fig:index}, this involves three components.

\textbf{Structural Entropy Correlation (SEC) Score.}
To select informative text pairs for clustering, \sys scores each pair of text nodes $(v_i, v_j)$ using a Structural Entropy Correlation (SEC) measure that combines semantic similarity with structural divergence~\cite{7456290}:
\begin{equation}\label{eq:sec}
  \mathrm{SEC}(v_i, v_j) \;=\; \alpha \cdot \cos(\mathbf{e}_i, \mathbf{e}_j)
    \;-\; (1-\alpha)\cdot\frac{\mathcal{H}^T(v_i, v_j)}{\log_2 |V|}
\end{equation}
where $\mathbf{e}_i$ is the embedding of node $v_i$, $\cos(\cdot,\cdot)$ denotes cosine similarity, and $\mathcal{H}^T(v_i, v_j) = -\sum_{v \in P_{ij}} \frac{d_v}{2|E|}\log_2 \frac{d_v}{\mathrm{vol}(T_v)}$ is the structural entropy along the tree path $P_{ij}$ between the two nodes, with $d_v$ the degree of $v$, $|E|$ the number of edges in the document tree, $\mathrm{vol}(T_v)$ the volume of the subtree rooted at $v$, and $|V|$ the total number of nodes. We set $\alpha{=}0.6$ and select the top-$k$ pairs with highest SEC scores as anchors for clustering.
Intuitively, SEC favors node pairs that are semantically similar yet structurally distant in the document tree---precisely the cross-section relationships that local retrieval misses.

\textbf{Pairwise-Constrained K-Means (PC-KMeans).}
Given the top-$k$ anchor pairs, an LLM labels each as must-link ($\mathcal{M}$) or cannot-link ($\mathcal{C}$). \sys then clusters all text nodes into $K$ clusters by minimizing the PC-KMeans objective~\cite{Basu2004ActiveSF}:
\begin{equation}\label{eq:pckm}
  \mathcal{J} = \sum_{l=1}^{K}\sum_{v_i \in C_l} \|\mathbf{e}_i - \boldsymbol{\mu}_l\|^2
    + w \!\!\sum_{(v_i,v_j)\in\mathcal{M}}\!\! \mathbb{1}[c_i{\neq}c_j]
    - w \!\!\sum_{(v_i,v_j)\in\mathcal{C}}\!\! \mathbb{1}[c_i{=}c_j]
\end{equation}
where $K$ is the number of clusters, $\boldsymbol{\mu}_l$ is the centroid of cluster $C_l$, $c_i$ denotes the cluster assignment of $v_i$, and $w$ is the penalty weight for constraint violations. By restricting LLM calls to the $k$ SEC-selected anchor pairs, \sys bounds annotation cost at $O(k)$.
The must-link and cannot-link constraints inject LLM-level semantic judgment into clustering, pulling semantically related nodes together even when their embeddings are not nearest neighbors.

\textbf{Hyper-Edges.}
Within each section, \sys uses LLM-based grouping to identify semantically related paragraphs.
Each group forms a hyper-edge connecting multiple text nodes, accompanied by a concise summary (30--50 words).
Hyper-edges residing in the same PC-KMeans cluster are linked via must-link relationships, creating a semantic overlay that spans the entire document.

\subsection{Filtering and Retrieval}
\label{sec:filtering}

Given a natural-language filter condition, \sys combines evidence from three complementary retrieval strategies: (i)~\emph{document-tree traversal}, which searches title nodes first and then retrieves content under the most relevant sections; (ii)~\emph{FAISS semantic search}~\cite{johnson2019billion}, which scores each node as $s(v_i, q) = w_{\tau(v_i)} \cdot \cos(\mathbf{e}_i, \mathbf{e}_q) \cdot \gamma^{d_i}$, where $w_{\tau}$ is a type weight (\eg title${=}1.5$, text${=}1.0$, table${=}0.8$), $d_i$ is the depth, and $\gamma \in (0,1)$ is a decay factor favoring shallower nodes; and (iii)~\emph{hyper-edge matching}, which searches hyper-edge summaries and retrieves all connected nodes for matching edges.
The combined evidence is passed to an LLM, which renders a boolean filtering decision per document.
For follow-up questions, the same retrieval pipeline supplies context passages to the LLM for RAG-based answer generation.

% ============================================================
\section{Demonstration}
\label{sec:demo}

% --- Figure 4: Example System Index View Screenshot Figure ---
\begin{figure}[t]
  \centering
  \includegraphics[page=6,
                    trim=5cm 0cm 5cm 0cm,
                    clip,
                    width=1.0\linewidth]
                    {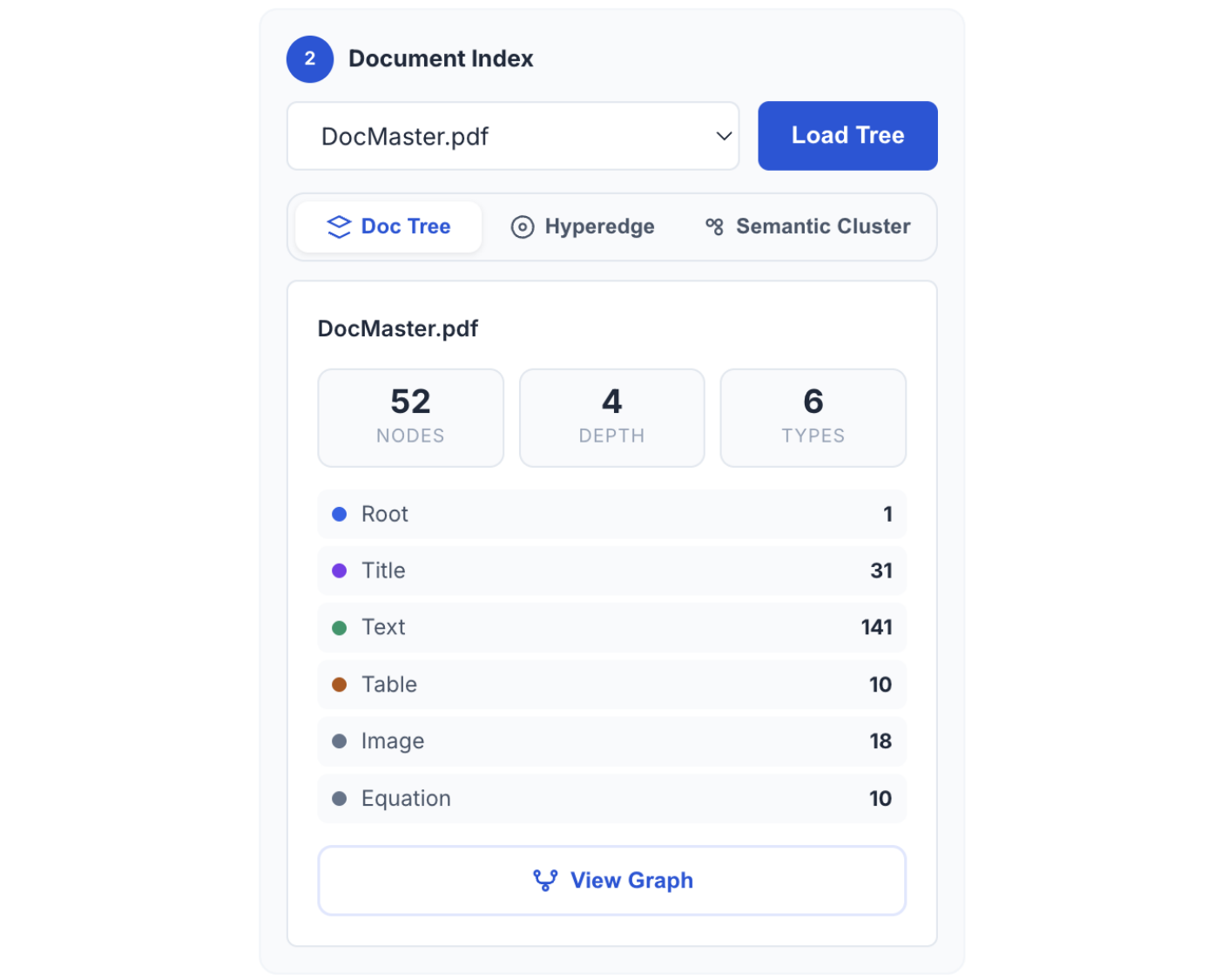} 
  \caption{Document tree view of \sys, showing the hierarchical structure of an example paper.}
  \Description{Document tree interface showing sections and content nodes from an example paper.}
  \label{fig:tree}
\end{figure}

% --- Figure 5: Example Document Statistics Screenshot Figure ---
\begin{figure}[t]
  \centering
  \includegraphics[page=8,
                    trim=1cm 0cm 1cm 0cm,
                    clip,
                    width=0.9\linewidth]
                    {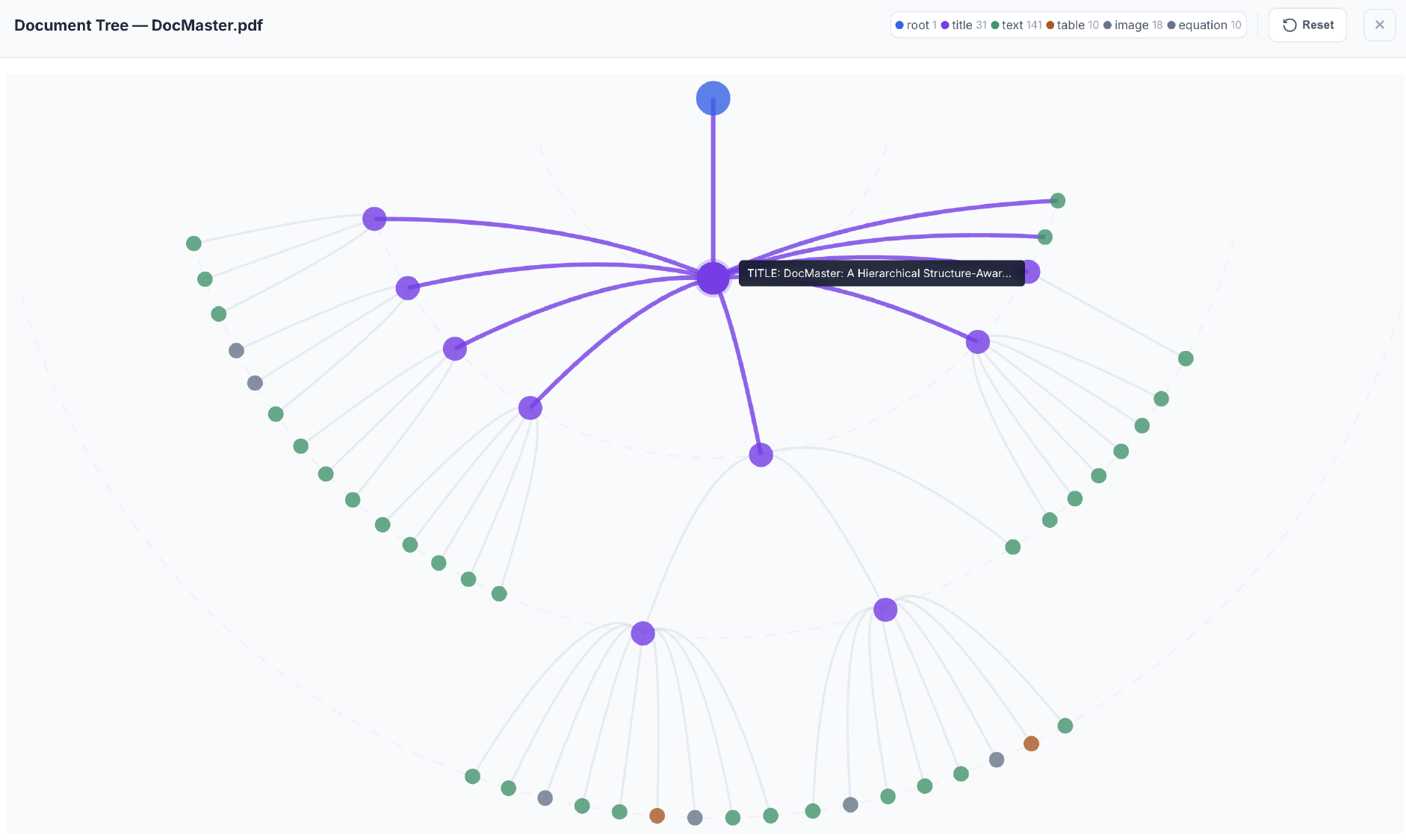} 
  \caption{Overview of Semantic Index.}
  \Description{Semantic index overview showing clustered document content and related index information.}
  \label{fig:index}
\end{figure}

\sys is deployed as a web application built with a React frontend and a FastAPI backend.
The demonstration showcases three interactive scenarios.
Figure~\ref{fig:system} shows the system frontend; Figure~\ref{fig:usecase} illustrates the end-to-end pipeline of \textbf{Scenario~3}; and Figures~\ref{fig:tree}--\ref{fig:index} display the document index views.

\textbf{Scenario 1: Document Upload and Exploration.}
A researcher uploads a folder of AI papers via drag-and-drop.
\sys processes each PDF in real time, progressively building the hierarchical document tree and semantic indices; a progress bar tracks the parsing status of each document.
The researcher then explores the document tree in a collapsible tree layout (Figure~\ref{fig:tree}), clicking any node to inspect its raw content, LLM-generated summary, and dense embedding.

\textbf{Scenario 2: Semantic Index Inspection.}
The interface provides three switchable views for inspecting the constructed indices (Figure~\ref{fig:index}): (i)~a \emph{Document Tree} view showing the structural hierarchy; (ii)~a \emph{Hyper-Edge} view displaying semantic groupings and must-link connections across sections; and (iii)~a \emph{Semantic Cluster} view visualizing PC-KMeans clustering results with adjustable hyperparameters ($\alpha$, $k$, $K$, $w$).
For instance, hyper-edges may link a ``retrieval mechanism'' paragraph in one paper's method section to an ``evaluation on open-domain QA'' paragraph in its experiments---a cross-section relationship invisible to flat retrieval.

\textbf{Scenario 3: Live Filtering and Follow-up Q\&A.}
The researcher enters a natural-language filter condition (\eg ``Does this paper propose a retrieval-augmented generation method?'') into the query panel.
\sys retrieves evidence from all three retrieval strategies, highlights matching passages in the tree, and displays a boolean filtering decision per document.
As shown in Figure~\ref{fig:usecase}, once relevant papers are identified, the researcher issues a follow-up question (\eg ``What retrieval strategies does this paper use?'') and receives a grounded answer with cited passages drawn from the filtered documents.
Users can also compare filtering results across different hyperparameter settings side by side (Figure~\ref{fig:system}).

% ============================================================
\section{Related Work}
\label{sec:related}

\textbf{Retrieval-Augmented Generation.}
RAG systems~\cite{lewis2020retrieval,borgeaud2022improving,zhou2025depth} ground LLM outputs in retrieved passages; Gao \etal~\cite{gao2023retrieval} provide a comprehensive survey.
Most systems treat documents as flat chunk sequences.
RAPTOR~\cite{sarthi2024raptor} builds hierarchical trees via recursive clustering, but purely from embedding similarity without preserving the original document hierarchy or modeling cross-section relationships.
% \sys preserves the author-intended structure and augments it with a semantic overlay capturing cross-section links.

\textbf{Document Parsing.}
MinerU~\cite{wang2024mineru} and Nougat~\cite{blecher2023nougat} extract structured content from PDFs; LayoutLM~\cite{xu2020layoutlm} learns layout-aware representations.
\sys builds on MinerU and extends it with hierarchical tree construction and semantic indexing.

\textbf{Constrained Clustering.}
Wagstaff \etal~\cite{10.5555/645530.655669} introduce constrained K-Means with instance-level supervision; Basu \etal~\cite{Basu2004ActiveSF} extend it to pairwise constraints.
\sys adapts PC-KMeans with LLM-generated constraints derived from document semantics, combining clustering efficiency with LLM judgment.

% ============================================================
\section{Conclusion}
\label{sec:conclusion}

We presented \sys, a hierarchical structure-aware document analysis system that leverages LLMs to analyze complex documents.
\sys first parses documents into hierarchical document trees preserving original layouts and constructs structure-aware semantic indices, then filters relevant documents from large collections via tri-modal retrieval, and finally supports follow-up question answering over the filtered results through RAG.
We demonstrated \sys through an interactive web interface covering the full workflow---from document-collection upload and semantic index construction to live filtering and question answering.
Future work includes building cross-document semantic indices that link related content across documents and supporting incremental index updates as collections evolve.

% ============================================================
\balance
\bibliographystyle{ACM-Reference-Format}

\bibliography{references}

\end{document}